\documentclass[%
 aip,
 amsmath,amssymb,
 reprint,%
]{revtex4-1}
\usepackage{xcolor}
\usepackage{float}
\usepackage{graphicx}
\usepackage{dcolumn}
\usepackage{bm}
\usepackage{makecell}
\usepackage[utf8]{inputenc}
\usepackage[T1]{fontenc}
\usepackage{mathptmx}
\usepackage{multirow}

\newcommand{\subb}[1]{$_{\mathrm{#1}}$}
\newcommand{\supp}[1]{$^{\mathrm{#1}}$}

\begin{document}

\preprint{AIP/123-QED}

\title[Double and triple ionisation of allene]{Single photon double and triple ionisation of allene}

\author{V. Ideböhn}
    \affiliation{Department of Physics, University of Gothenburg, Origov\"agen 6B, 412 58 Gothenburg, Sweden}
\author{A.J. Sterling}
    \affiliation{Department of Chemistry, Chemistry Research Laboratory, Mansfield Road, Oxford OX1 3TA, United Kingdom}%
\author{M. Wallner}%
    \affiliation{Department of Physics, University of Gothenburg, Origov\"agen 6B, 412 58 Gothenburg, Sweden}
\author{E. Olsson}
    \affiliation{Department of Physics, University of Gothenburg, Origov\"agen 6B, 412 58 Gothenburg, Sweden}
\author{R.J. Squibb}
    \affiliation{Department of Physics, University of Gothenburg, Origov\"agen 6B, 412 58 Gothenburg, Sweden}
\author{U. Miniotaité}
    \affiliation{Department of Physics, Chalmers University of Technology, Kemigården 1, 412 96 Gothenburg, Sweden}
\author{E. Forsmalm}
    \affiliation{Department of Physics, University of Gothenburg, Origov\"agen 6B, 412 58 Gothenburg, Sweden}
\author{M. Forsmalm}
    \affiliation{Department of Physics, University of Gothenburg, Origov\"agen 6B, 412 58 Gothenburg, Sweden}
\author{S. Stranges}
    \affiliation{IOM-CNR Tasc, SS-14, Km 163.5 Area Science Park, Basovizza 34149, Trieste, Italy}%
    \affiliation{Dipartimento di Chimica e Tecnologie del Farmaco, Universitá Sapienza, Rome I-00185, Italy}%
\author{J.M. Dyke}
    \affiliation{School of Chemistry, University of Southampton, Highfield, Southampton SO17 1BJ, United Kingdom}%
\author{F. Duarte}
    \affiliation{Department of Chemistry, Chemistry Research Laboratory, Mansfield Road, Oxford OX1 3TA, United Kingdom}%
\author{J.H.D. Eland}
\affiliation{Department of Chemistry, Physical and Theoretical Chemistry Laboratory, Oxford University, South Parks Road, Oxford OX1 3QZ, United Kingdom} 
\author{R. Feifel}
    \email{raimund.feifel@physics.gu.se}
    \affiliation{Department of Physics, University of Gothenburg, Origov\"agen 6B, 412 58 Gothenburg, Sweden}

\date{\today}

\begin{abstract}
Double and triple ionization of allene are investigated using electron-electron, ion-ion, electron-electron-ion and electron-electron-ion-ion (ee, ii, eei, eeii) coincidence spectroscopies at selected photon energies. The results provide supporting evidence for a previously proposed roaming mechanism in H$_3^+$ formation by double ionisation. The lowest vertical double ionization energy is found to be 28.5 eV, while adiabatic double ionisation is not accessed by vertical ionisation at the neutral geometry. The triple ionization energy is found to be close to 50 eV in agreement with theoretical predictions. The doubly charged parent ion is stable up to about 2 eV above threshold, after which dissociations by charge separation and by double charge retention occur with comparable intensities. Fragmentation to H$^+$ + C$_3$H$_3^+$ starts immediately above threshold as a slow (metastable) decay with 130±10 ns mean lifetime.
\end{abstract}

\maketitle

\begin{quotation}
%
\end{quotation}


\section{\label{sec:level1}Introduction}

Allene (propadiene, $\mathrm{C_3H_4}$, CH2:C:CH2) is the prototype and first member of the group of molecules called cumulenes.  It is thought to be present in the interstellar medium \cite{kaiser1997neutral, jones2011formation} at least as an intermediate, but as it lacks a dipole moment it has not yet been detected there directly.  It is closely related to the important transient molecule ketene (CH2:C:CO) which has been detected in the interstellar medium \cite{ruiterkamp2007organic}.  This paper will report on single-photon double and triple ionisation of allene, as well as on allene’s dissociation pathways.

The single ionisation of allene and its consequent dissociations have been studied using mass spectrometry \cite{stockbauer1979ionization}, photoelectron spectroscopy \cite{baker1969photoelectron, thomas1974photoelectron,bieri1977valence} and photoelectron-photoion coincidence spectroscopy \cite{dannacher1978unimolecular}.  The electronic structure of the singly charged molecule has recently attracted theoretical interest, focusing on “Möbius” and “helical” frontier orbitals \cite{garner2018coarctate, hendon2013helical,soriano2014allenes} peculiar to cumulenes. 

Much less is known about multiple ionisation of allene, but double and triple ionisation of the molecule  have both been examined in experiments using the electron capture charge exchange technique, \cite{andrews1992allene,andrews1995allene}, where the lowest vertical double ionization energy to the expected ground-state triplet was determined as 28.2 $\pm 0.3$ eV. Allene double ionisation and the rates of its dissociation reactions have also been examined using fs IR multi-photon laser pulses \cite{hoshina2011metastable,xu2013experimental}.  In these and other previous studies of allene double ionization, dissociation producing H$_3^+$ ions has been found, surprisingly, to be more abundant than from molecules where three hydrogen atoms are initially contiguous in  – CH$_3$ groups. The mechanism was interpreted in the specific case of H$_3^+$ formation from a series of hydrocarbon molecules with no methyl group including allene as involving a “roaming” mechanism where a neutral H$_2$ is detached from the residual C$_3$H$_2^{2+}$ fragment, after which a proton is transferred from the doubly charged parent to form the H$_3^+$ ion \cite{hoshina2008publisher}, in line with the theoretical work of Mebel and Bandrauk \cite{mebel2008theoretical}. The creation of H$_3^+$ is found to be a significant dissociation mechanisms for doubly charged allene, and  must include at least migration of one hydrogen atom \cite{hoshina2011metastable, mebel2008theoretical} and possibly more extensive rearrangement. A number of experiments probing this mechanism, mainly by IR multi-photon ionisation in different molecules (mostly methanol) followed \cite{livshits2020time}. Calculations on methanol indicate that the formation of H$_3^+$ competes with that of H$_2^+$ on a sub-100 fs time scale \cite{livshits2020time}. 
In the present work we examine both H$_2^+$ and H$_3^+$ detection in correlation with the electrons involved in the creation of the nascent allene dications which dissociate to form them.

Our study was carried out using electron-electron (ee) coincidence measurements at photon energies above and below the C1s inner shell (approximately 291 eV), and electron-electron-ion (eei) as well as electron-electron-ion-ion (eeii) coincidence measurements at 40.8 eV. These measurements determine the spectra of states of nascent doubly ionised allene, and the spectra coincident with undissociated parent dications as well as each set of dissociation products.  Photoionization mass spectra were also measured at all photon energies and ion-ion coincidence maps were taken to identify the decay pathways and clarify the mechanisms. The spectra and dissociation pathways are discussed and interpreted with the help of quantum chemical calculations carried out by ourselves and are also compared with calculations available in the literature \cite{mebel2008theoretical}. 

\section{Experimental Methods}

The experiments were carried out in our laboratory at the University of Gothenburg and at the synchrotron radiation facility BESSY-II of the Helmholtz Zentrum für Materialien in Berlin. In both cases, the target gas was let into the spectrometer using a hollow needle to create an effusive gas beam in the interaction region. The basic configuration of our system, which builds on a more compact version of the original instrument of this type \cite{eland2003}, comprises a strong conical magnet with a divergent field of approximately 1 T at the light-matter interaction point and a 2 m flight tube surrounded by a weak homogeneous solenoid field (few mT). At the end of the flight tube, electrons are registered by a multi-channel plate (MCP) detector, making the overall collection-detection efficiency of this magnetic bottle electron spectrometer (MBES) for low energy (< ca. 400 eV) electrons about 50-60 \%. The electron energy resolution of the set-up is about $\mathrm{E_{kin}/\Delta E_{kin} \sim 50}$ when collecting only electrons. For collection of both electrons and ions the strong conical magnet is replaced by a hollow ring magnet with a weaker magnetic field in the interaction zone, which limits the electron energy resolution to about $\mathrm{E_{kin}/\Delta E_{kin} \sim 20}$, but with the benefit that ions can be extracted in the opposite direction to the electrons \cite{eland2006,feifel2006}. In this latter configuration, electron-ion coincidence data are obtained by first letting the electrons leave the interaction region, before accelerating the ions in the opposite direction, through the ring magnet and into a two-field time of flight mass spectrometer with a 0.12 m long flight tube. The fields are optimized to achieve time focus conditions, giving a numerical resolution for thermal ions of about 25. Under these conditions peaks in the parent ion group are only partially resolved. Complementary non-pulsed ion-only measurements were carried out with the magnetic bottle in its basic configuration by using the 2 m long flight tube for ions instead of electrons, and similar measurements were made using the shorter flight tube to capture processes on shorter time scales. With the longer flight path the numerical mass resolution was about 50, sufficient to resolve all the main ion peaks. 

In the Gothenburg laboratory, a pulsed helium gas discharge lamp with a repetition rate of approximately 4 kHz was used as ionisation source where the atomic emission lines of HeI$\alpha$ and HeII$\alpha$ provided photon energies of 21.2 and 40.8 eV, respectively. The discrete energies were selected using a monochromator based on an ellipsoidal grating of 550 lines/mm groove density which also focuses the radiation of a selected wavelength into the interaction region. 

The flight times of the electrons were converted to kinetic energies on the basis of a calibration derived from measurements of known photoelectron and Auger electron spectra. For calibration in the low energy region we used the spectra of molecular oxygen at 21.2 and 40.8 eV photon energy, and for higher energies the spectra of atomic argon and krypton at photon energies of 100 eV and above. 

At BESSY-II, the experimental set-up was mounted on undulator beamline UE52/SGM where the photon energy can be tuned in the soft X-ray energy region. In order to doubly ionize allene through valence-valence, core-Auger and core-valence electron removal photon energies of 100 eV, 110 eV, 300.5 eV and 350.4 eV were used. The set up was essentially the same as in Gothenburg but with the addition of a mechanical chopper synchronized to the radio frequency signal of the storage ring operating in single bunch mode. The chopper was set to increase the time interval between photon bunches passed to the experiment (otherwise 805.5 ns) to between 10 and 100 $\mu$s, to allow detection of electrons and ions without interference from subsequent radiation pulses during their expected maximum flight times.
The stated purity of the sample was 99.9 \%. This was verified by on-line valence and core level photoelectron spectroscopy and mass spectroscopy which showed no impurity lines.

\section{Theoretical Methods}

All calculations were carried out using the ORCA program (v 4.1.2) \cite{neese18}. Optimisations were carried out at the CASSCF(4,4) and CASSCF(2,4) levels with the def2-TZVP basis set \cite{weigend05} for neutral and doubly-ionised states respectively, reflecting ionisation from the 1e and 2e ($\pi$) orbitals of allene. Ionisation energies were obtained at the NEVPT2 level \cite{angeli01}, based on CASSCF(4,4) and CASSCF(2,4) wavefunctions for neutral and doubly-ionised states respectively. NEVPT2 calculations used the cc-pVQZ basis set \cite{dunning89}.
Static correlation arising from (near)-degeneracy, that is often a challenge for single-reference methods \cite{andrews1992allene}, is explicitly accounted for using the current procedure, allowing the accurate calculation of the double ionisation energy to the S$_0$ state. There is good agreement on the vertical double ionisation energy with the extensive calculations of Mebel and Bandrauk\cite{mebel2008theoretical}, who also found that after geometry optimization i.e. relaxation, the lowest singlet state becomes much lower in energy than the triplet.

\section{Results and Discussion}

\subsection{Double ionisation of Allene}

\subsubsection{Valence double ionisation}

\begin{figure}[H]
\centering
\includegraphics[width = 0.5\textwidth]{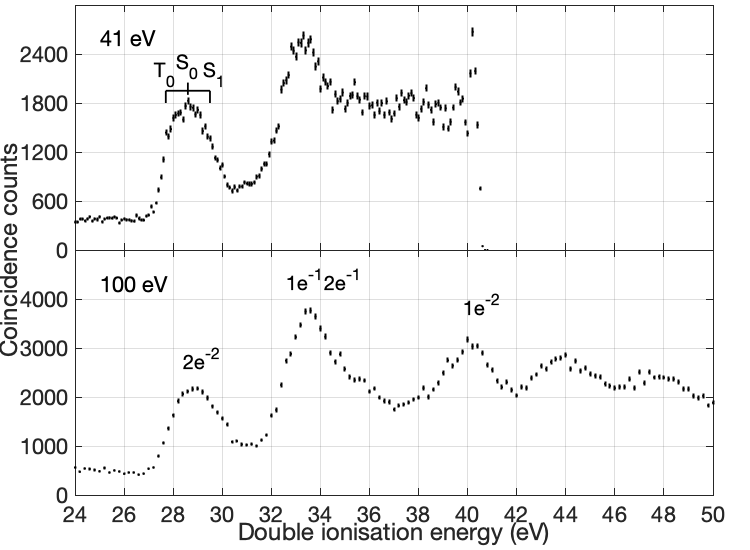}
\caption{\label{fig:DIP} Double ionisation electron spectra of allene measured at the photon energies of 40.8 eV (upper panel) and 100 eV (lower panel). The 40.8 eV data were taken in Gothenburg with a helium gas discharge lamp and the 100 eV data were taken at BESSY-II in Berlin. Both spectra are based on electron pair measurements. Assignments in terms of leading configuration (lower panel), and for the lowest structure in the 40.8 eV spectrum in terms of different electronic states (upper panel) are included.}
\end{figure}

Figure \ref{fig:DIP} shows double ionization spectra from electron-electron coincidence measurements at photon energies of 40.8 and 100 eV. Accidental coincidences have been subtracted, but a constant background remains. 
The two spectra agree well with each other, reflecting the onset of double ionization at about 27 eV as part of the first band with a vertical double ionisation energy of about 28.5 eV, the latter in good agreement with the theoretical value of 28.05 eV \cite{mebel2008theoretical}. Additional structures appear above 30 eV double ionisation energy, which are centred near 33.5, 36, 40, 44 and 48 eV and which involve removing more tightly bound electrons in allene. The bands associated with electron removal from the degenerate orbitals are expected to be the most intense because of the larger number of electrons in them, but they probably conceal underlying bands from combinations with the other ($\sigma$-type) orbitals.

\begin{figure}[H]
\centering
\includegraphics[width = 0.5\textwidth]{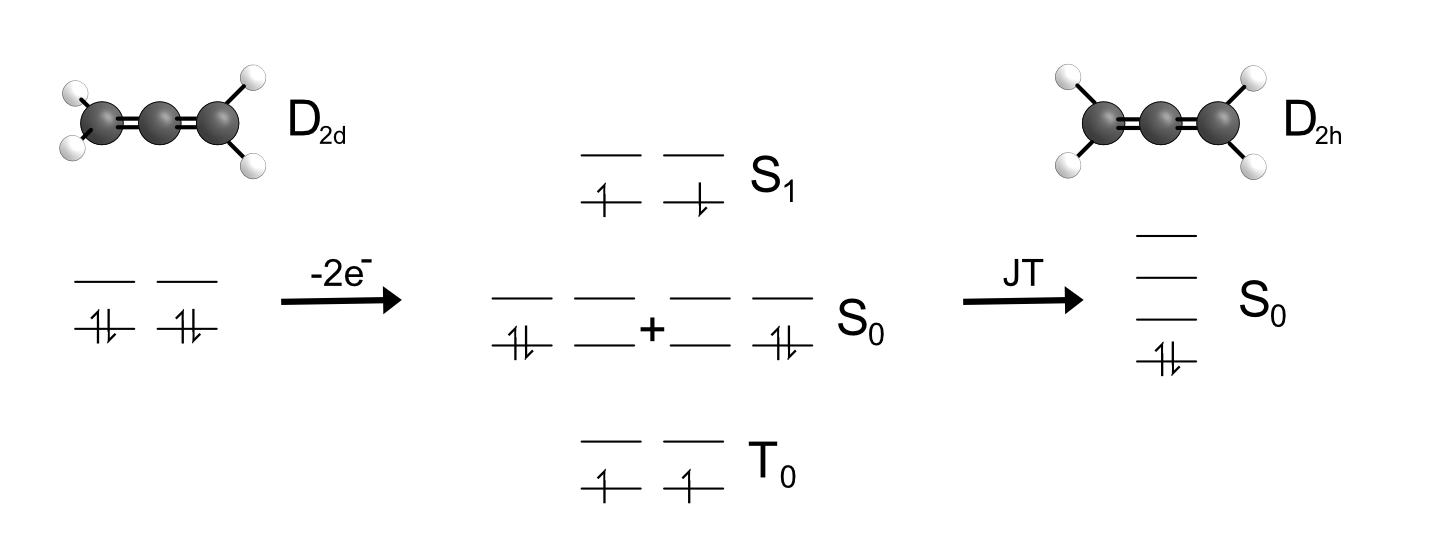}
\caption{\label{fig:DIP_structure} Changes in the  orbital structure of neutral allene in transitions to the three states that make up the first peak in the double ionisation spectrum in Fig. \ref{fig:DIP}. Neutral allene is stripped of two electrons in three different ways, ending up in three different states. The S$_0$ state rearranges by Jahn-Teller distortion, resolving the degeneracy of two equivalent forms of the state in D$_{2d}$  symmetry, resulting in a planar molecule described by the D$_{2\mathrm{h}}$-point group instead.} 
\end{figure}

The first band is especially interesting, with substructures at 27.7, 28.6 and 29.5 eV, which are interpreted with the aid of our calculations of vertical ionisation energies to the lowest triplet T$_0$ (\supp{3}A\subb{2} in D\subb{2d}) state, the lowest singlet S\subb{0} state, which undergoes Jahn-Teller distortion, possibly to \supp{1}A\subb{g} in D\subb{2h} or to a lower symmetry structure, and the second singlet S\subb{1} state which retains D\subb{2d} geometry, as illustrated in Fig. \ref{fig:DIP_structure}. 

Our theoretical values for the double ionisation energy of the different states as well as the experimental values taken from Fig. \ref{fig:DIP}, are summarized in Table \ref{tab:DIP}. The theoretical and experimental values are very close to each other, but adiabatic ionisation to the rearranged ground state is not observed. Our calculations are in excellent agreement with the double ionisation calculations by Andrews et al. \cite{andrews1992allene, andrews1995allene} at the MP2 and MP4 levels of theory. In the double ionisation of allene, the molecule goes from being described by the point group D$_{2d}$ to D$_{2h}$ through a reorganisation of the orbitals and a reduction of degenerate states. This process, where the 3D structure of neutral allene changes to a planar form upon double ionisation is schematically illustrated in Fig. \ref{fig:DIP_structure}. Although their calculations are otherwise very extensive, Mebel and Bandrauk \cite{mebel2008theoretical} did not report calculated energies or structures for S$_1$ or any higher states of doubly ionised allene.


 The lowest (adiabatic) double ionisation energy is calculated as 26.1 eV, well below the range seen to be accessed by single-photon ionisation in our spectra. The triplet state also relaxes, but much less, ending up at an adiabatic double ionisation energy of 27.6 eV, which is just in the range seen in our spectrum.

\begin{table}[H]
    \centering
    \caption{Theoretical values for the vertical and 0-0 double ionisation of allene from calculations carried out here using CASSCF(4,4) and CASSCF(2,4) for neutral and doubly-ionised states, respectively. For comparison, experimental values extracted from Fig. \ref{fig:DIP} are also included.}
    \begin{tabular}{|c|c|c|c|}
    \hline
    \textbf{\makecell{Electronic \\ state}} & \textbf{\makecell{Vertical double \\ ionisation (eV)}} & \textbf{\makecell{0-0 double \\ ionisation (eV)}} & \textbf{\makecell{Experiment \\ (eV)}} \\
    \hline
    S$_0$ & 28.6 & 26.1 & 28.6 \\
    \hline
    S$_1$ & 29.5 & 29.2 & 29.5 \\
    \hline
    T$_0$ & 28.2 & 28.0 &  27.7 \\
    \hline
    \end{tabular}
    \label{tab:DIP}
\end{table}

\subsubsection{Double ionisation by Auger decay}

\begin{figure}
\centering
\includegraphics[width = 0.5\textwidth]{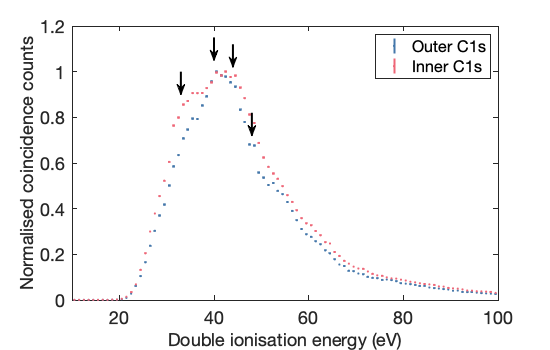}
\caption{\label{fig:DIP_Auger} Double ionisation spectra of allene measured with 300.5 eV photon energy. The spectra have been selected on outer and inner C1s core electron, respectively. The inner C1s electron has a binding energy of 290.8 eV and the outer carbon 1s electron has 290.6 eV \cite{travnikova2008structure}, and the selections comprise binding energy ranges of $\pm 0.1$ eV. 
The arrows indicate structures in the ionization spectra at energies of about 33, 40, 44 and 48 eV.}
\end{figure}

Electronic states with two vacancies in valence orbitals can also be formed by initial inner shell (C1s) ionization followed by emission of a secondary Auger electron upon filling of the short-lived core hole. In the case of allene, single ionisation from the C1s orbitals was reported by Travnikova et al. \cite{travnikova2008structure} to occur at 290.6 and 290.8 eV, giving rise to photoelectron lines that were only partially resolved in our electron pair measurements carried out at 300.5 eV photon energy. In analyzing the coincidence data, partially separate selection is possible by choosing the extreme sides of the composite photoelectron line feature.  Double ionisation electron spectra (incorporating the energy of the selected photoelectrons) obtained in this way are shown in Fig. \ref{fig:DIP_Auger}.

As can be seen, there is a clear difference between the two selectively extracted spectra. Also, even though the experimental resolution is low (ca. 5 eV) because of the high electron kinetic energies, substructures that correspond well with the distinct bands seen in the 100 eV spectra (cf. Fig. \ref{fig:DIP}) are discernible as indicated by arrows, at about 33, 40, 44 and 48 eV.

\subsection{Triple ionisation of Allene}

\begin{figure}
\centering
\includegraphics[width = 0.5\textwidth]{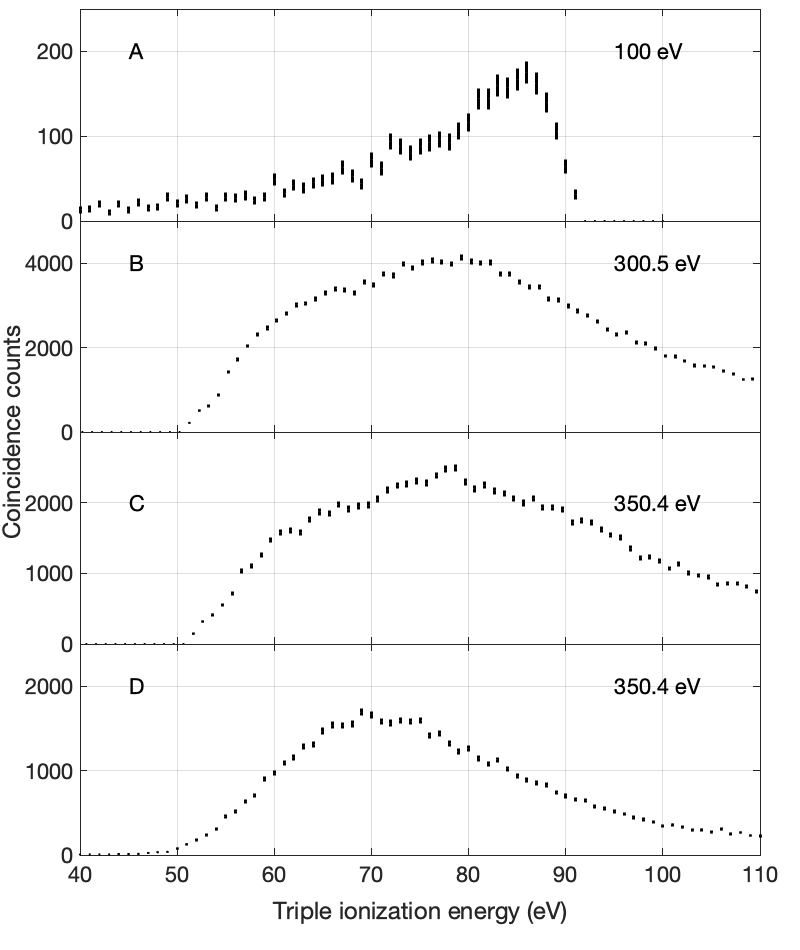}
\caption{\label{fig:TIP} Triple ionization at 100, 300.5 and 350.4 eV photon energies. In panel A, the triple ionization spectrum has been extracted by limiting the data to include events with 100 eV photons where the electron with the lowest energy is 3$\pm$0.5 eV. In panels B and C, events with one core electron emission and two other electron emissions have been selected. The triple ionization spectrum in panel D is derived by identifying events where a core-valence doubly ionised state is formed before an Auger electron is emitted.}
\end{figure}

Triple ionization of allene can occur by different main pathways, according to the photon energy.  At 100 eV there is only triple valence electron removal. At all energies above about 300 eV core ionization and subsequent double Auger decay is dominant, while at and above 330 eV there is also core-valence double ionization followed by Auger decay. Spectra corresponding to the three different pathways are shown in Fig. \ref{fig:TIP}. 

The 100 eV spectrum (panel A) is essentially structureless, showing only a smoothly rising signal with a possible start at about 50 eV (and an uncertainty of nearly 2 eV). To remove spurious low energy electrons, the lowest energy included in the events was selected to 3$\pm$0.5 eV. 

The triple ionization spectra displayed in panels B and C were extracted from measurements at 300.5 and 350.4 eV photon energy by selecting events with one core electron and two other electrons. These selections were made by visual inspection of the electron-electron coincidence maps \textcolor{blue}{(?)} where electrons that were not part of the desired energy sharing have been removed. As can be seen, the spectra reflect a broad, featureless bump with onset at about 51 ± 2.3 eV and maximum intensity in the 79 – 80 eV energy range. We note that the ratio of double to single Auger is experimentally found to be 13\% at both 300.5 eV and 350.4 eV photon energy which is slightly higher than reported ratios in other C\subb{3}  carbon compounds from the work of Hult Roos et al. \cite{hult2019multi}. 

At 350.4 eV we could also extract the triple ionization spectrum shown in panel D, based on selection of initial core-valence double ionization, for which full spectra will be presented in a forthcoming publication, followed by an additional Auger electron emission. The spectrum locates the lowest triple ionization energy at about 50 $\pm$ 5.4 eV. All these onset values are in reasonable agreement with the calculated adiabatic triple ionisation values given by Mebel and Bandrauk \cite{mebel2008theoretical} as 52.29 eV, and 53.05 eV from different forms of theory. 

\subsection{Fate of Allene above lowest double ionisation threshold}

To investigate the fate of allene exposed to photon energies above the lowest double ionisation threshold ion detection in multiplex was employed.

\subsubsection{Ion time-of-flight mass spectra}

\begin{figure}[h!]
\centering
\includegraphics[width = 0.5\textwidth]{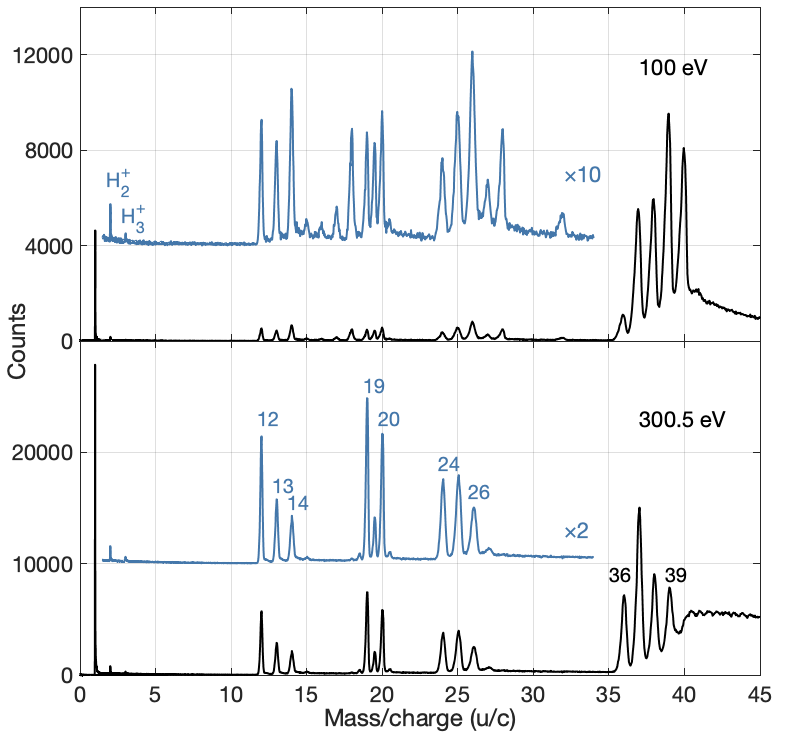}
\caption{\label{fig:mass_spectrum_100_301eV} Mass spectra of allene at photon energies 100 eV (upper panel) and 300.5 eV (lower panel). The mass spectrum at 40.8 eV (not shown) is very similar to the 100 eV spectrum. 
}
\end{figure}

Figure \ref{fig:mass_spectrum_100_301eV} shows mass spectra of allene obtained at the photon energies of 100 eV (upper panel) and 300.5 eV (lower panel). Both spectra were obtained using the 2 m flight tube for ion detection (instead of electron detection). The 100 eV spectrum is very similar to the mass spectrum at 40.8 eV which is not shown separately.

As can be seen, the 100 eV spectrum is dominated by the parent ion group (m/e = 40, 39, 38, 37, 36). The weak and otherwise unexpected peak(s) at 32 (and 28) suggest(s) that a small amount of air was present during this recording. More interestingly, there are relatively strong peaks for doubly charged ions containing the C$_3$ species at m/e = 20, 19.5 and 19, all three of roughly equal intensity. The very weak peak at m/e = 16 is probably due to O$^+$, while the peak at m/e = 15 of similar intensity is attributed to the CH$_3^+$ species. Its intensity is about 1/4 that of the m/e = 14 peak associated with CH$_2^+$. Very interestingly, there are distinct peaks for m/e = 2 and 3, i.e. for H\subb{2}\supp{+} and H\subb{3}\supp{+}.

The most striking points in comparing the mass spectrum at 300.5 eV with the 100 eV data are the greatly increased dissociation at 300.5 eV, where double ionisation by the Auger effect dominates, and the stability of the doubly-charged ions C$_3$H$_4^{2+}$, C$_3$H$_3^{2+}$ and C$_3$H$_2^{2+}$ (but not C$_3$H$^{2+}$ or C$_3^{2+}$). We note that the structure and energetics of the parent doubly charged ion were calculated by Mebel and Bandrauk  \cite{mebel2008theoretical}, but those of C$_3$H$_2^{2+}$ and C$_3$H$^{2+}$ were not. They, and the apparently less stable or unstable C$_3$H$^{2+}$ and C$_3^{2+}$ species are fundamental entities, likely to be of relevance in astrophysical contexts. From a theoretical point of view, not much is known about their structure, except for the C$_3^{2+}$ species for which a linear configuration has been suggested \cite{hogreve1995ab}. However, for C$_3^+$ there is experimental evidence from Coulomb explosion imaging \cite{faibis1987geometrical} and theory \cite{taylor1991ab,diaz2006ionization} that it is non-linear.

\subsubsection{Ion-ion coincidences}

\begin{figure}[h!]
\centering
\includegraphics[width = 0.5\textwidth]{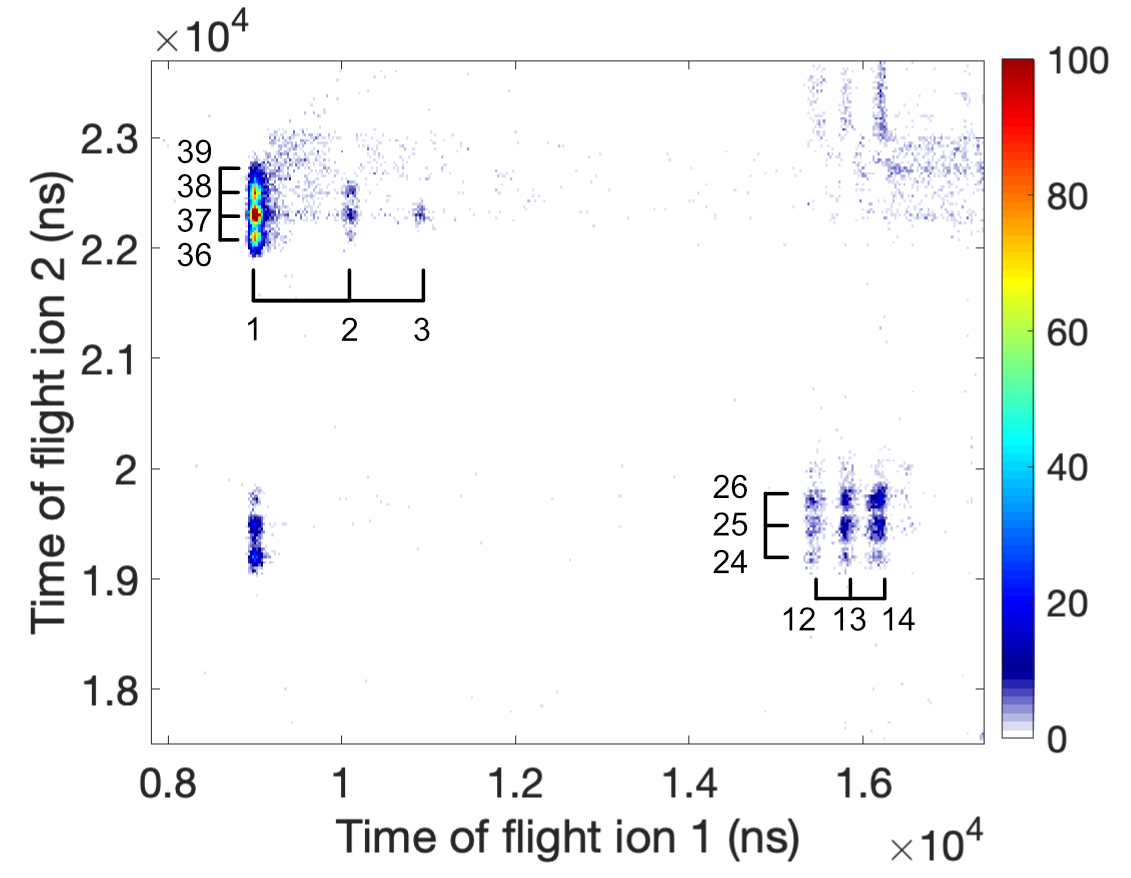}
\caption{\label{fig:ion_map_100}Ion-ion coincidence map at 100 eV photon energy using the 2 m flight tube. Accidental coincidences have been removed and the mass over charge information is included in the plot.}
\end{figure}

An ion-ion coincidence map associated with the 100 eV ion time-of-flight spectrum from Fig. \ref{fig:mass_spectrum_100_301eV} and after subtraction of accidental coincidences is shown in Fig. \ref{fig:ion_map_100}. The detectable ion pairs seen in this map are summarized in Table \ref{tab:ion_coincidences}.

\begin{table}[H]
    \centering
    \caption{\label{tab:ion_coincidences}Ion pairs detected in the 100 eV ion-ion coincidence map shown in Fig. \ref{fig:ion_map_100}. Note: the H\supp{+} + CH\supp{+} or C\supp{+} channels are not included in the figure.}
    \begin{tabular}{|l|l|}
    \hline
        \textbf{Ion pair} & \textbf{\makecell{Relative abundance \\ ($\Sigma = 1000$)} } \\
        \hline
         H\supp{+} + C\subb{3}H\subb{3}\supp{+} or C\subb{3}H\subb{2}\supp{+} or C\subb{3}H\supp{+} or C\subb{3}\supp{+} & 562 \\
         \hline
         H\supp{+} + C\subb{2}H\supp{+} or C\subb{2}\supp{+} & 92 \\
         \hline
         H\supp{+} + CH\supp{+} or C\supp{+}\supp{*} & 72 \\
         \hline
         H\subb{2}\supp{+} + C\subb{3}H\subb{2}\supp{+} or C\subb{3}H\supp{+} or C\subb{3}\supp{+} & 33 \\
         \hline
          H\subb{3}\supp{+} + C\subb{3}H\supp{+} or C\subb{3}\supp{+} & 12 \\
          \hline
          C\supp{+} + C\subb{2}H\subb{3}\supp{+} or C\subb{2}H\subb{2}\supp{+} or C\subb{2}H\supp{+} or C\subb{2}\supp{+} & 49 \\
          \hline
          CH\supp{+} + C\subb{2}H\subb{3}\supp{+} or C\subb{2}H\subb{2}\supp{+} or C\subb{2}H\supp{+} or C\subb{2}\supp{+} & 76 \\
          \hline
          CH\subb{2}\supp{+} + C\subb{2}H\subb{3}\supp{+} or C\subb{2}H\subb{2}\supp{+} or C\subb{2}H\supp{+} or C\subb{2}\supp{+} & 104 \\
          \hline
    \end{tabular}
\end{table}

Apart from the numerous ion pairs observed at 100 eV, some of which may come from triple ionisation and involve an unobserved third ion, four double-charge-retaining dissociation channels producing C\subb{3}H\subb{3}\supp{2+}, C\subb{3}H\subb{2}\supp{2+}, C\subb{3}H\supp{2+} and C\subb{3}\supp{2+} as seen in Fig. 5, arise entirely from double ionisation. From these and other data we identify the H\supp{+} + C\subb{3}H\supp{+} + H\subb{2} double ionisation channel as the most probable charge separation channel 

\begin{figure}[h!]
\centering
\includegraphics[width = 0.5\textwidth]{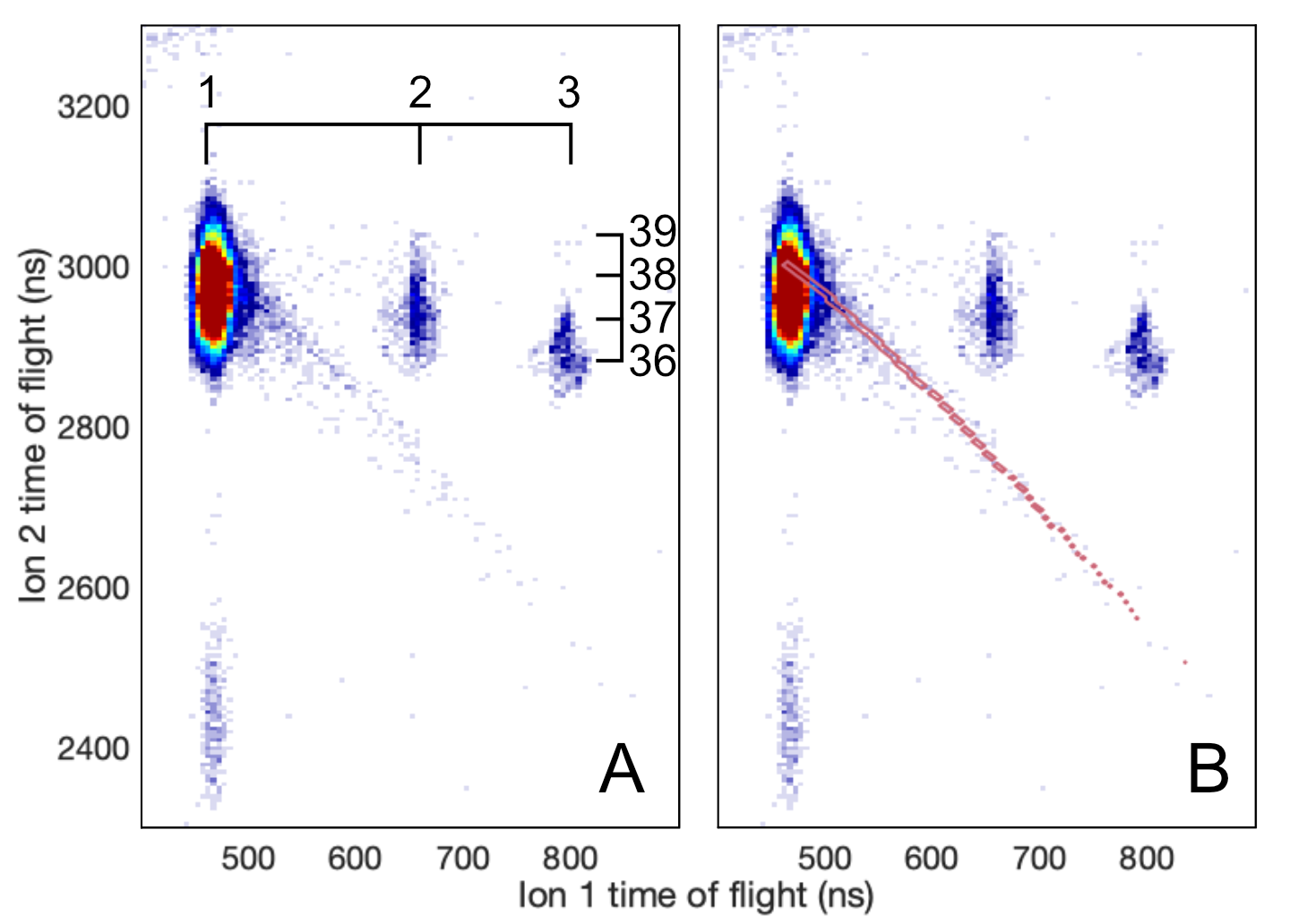}
\caption{\label{fig:metastable} Panel A displays an ion-ion coincidence map at 40.8 eV photon energy using the shorter (0.12 m) flight tube. Mass/charge is denoted by the numbers in the figure. The sloping feature which starts at the correlation island of mass 1 and 39 and extends towards the lower right corner, indicates the presence of a metastable doubly charged state with a lifetime on the order of the flight times of the ions in this shorter tube. This metastable state dissociates into the two singly charged ions seen in the map, H\supp{+} and C\subb{3}H\subb{3}\supp{+}, as illustrated in panel B by the pink markings. The markings have been obtained by simulating the decay in SIMION with a lifetime of the metastable state of 129$\pm$9.9 ns.}
\end{figure}

Fig. \ref{fig:metastable} displays an ion-ion coincidence map of allene obtained at the photon energy of 40.8 eV in DC mode using a much shorter (about 0.12 m long) ion time-of-flight spectrometer mounted in tandem configuration to the electron flight tube. This map confirms, though at lower resolution, several of the ion pair channels involving the H\supp{+}, H\subb{2}\supp{+}, and H\subb{3}\supp{+} species. The significant appearance of H\subb{3}\supp{+} ions in coincidence is in agreement with the findings of Hoshina et al. \cite{hoshina2011metastable} and related theory \cite{mebel2008theoretical}.  While the mechanism of formation of the H\subb{3}\supp{+} ions is discussed in the literature in terms of a “roaming” mechanism, which involves an initial H\subb{2} species that “orbits” within a doubly-charged precursor until it succeeds extracting the third proton, the identity of the electronic state from which it happens in allene has not been determined experimentally. 
\textcolor{magenta}{According to the energy range it could be any of the three states in the first double ionisation band (cf. Fig. \ref{fig:DIP}). If Mebel and Bandrauk’s RRKM rate calculation \cite{mebel2008theoretical} is taken at face value, it means that H$^+$ + C$_3$H$_3^+$ come from the relaxed ground state S$_0$. Further support for that can be obtained from figure 3 of Mebel and Bandrauk\cite{mebel2008theoretical} according to which the ground state can make hiving off "roaming H$_2$" more likely by the involvement of a methyl acetylene transition state which posses favourable structural characteristics.}


The most interesting aspect in the 40.8 eV map is the presence of an unambiguous, though weak metastable tail, which is essentially invisible at 100 eV and higher photon energies. The map identifies the most intense part of the tail as belonging to the H\supp{+} + C\subb{3}H\subb{3}\supp{+} reaction channel but its comparatively broad spread may include a contribution from the H\supp{+} + C\subb{3}H\subb{2}\supp{+}, too. The H\supp{+} + c-C\subb{3}H\subb{3}\supp{+} (cyclic cyclo-propenyl) ion pair has the lowest thermodynamic formation limit at 25 eV, while almost all the other observed pairs have limits near 26 eV for formation with no kinetic or internal energy release. Since there is no detectable “V” shape from the metastable centred on the doubly charged parent ion’s position as apex on the diagonal (t$_1$ = t$_2$) \cite{field1993lifetimes}, no longer-lived group of ions dissociates in the flight tube. This implies that essentially all the metastable decay takes place in the source field of our spectrometer. 

\begin{figure}[h!]
\centering
\includegraphics[width = 0.5\textwidth]{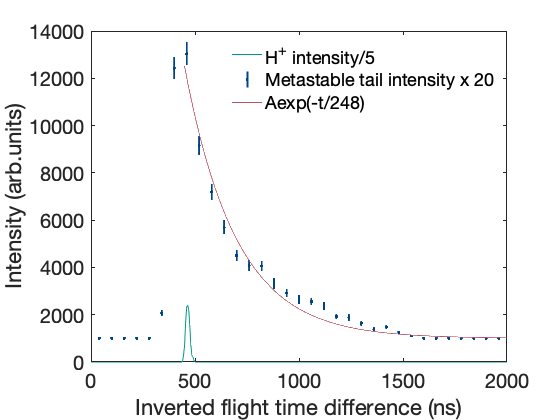}
\caption{\label{fig:lifetime} Metastable tail intensity as a function of the inverted flight time difference with a fit to the exponential decay in yellow. The coincidences are the same as in the tail in Fig. \ref{fig:metastable} plotted against t$_2$-t$_1$ where t$_1$ and t$_2$ are the flight times of the first and second ion to be detected, respectively. The location of the mass peak for H$^+$ ions on the same scale is included below the decay curve for comparison.}
\end{figure}

The existence of this metastable decay was already noted by Barber and Jennings in 1969 \cite{barber1969kinetic} in electron impact mass spectrometry. To investigate the metastable lifetime, we plotted the coincidence data as a map of t$_1$+t$_2$ vs. t$_2$-t$_1$, where the metastable tail gets concentrated as a strip almost parallel to the (t$_2$-t$_1$)-axis. The intensity as a function of the time difference was extracted and plotted, as is done in Fig. \ref{fig:lifetime}, for comparison with the theory given by Field and Eland \cite{field1993lifetimes}. The fit to the metastable state in the figure has a 95\% confidence interval of 229.2 - 266.8 ns which translates to a lifetime of 130.5 $\pm$ 9.9 ns. The observation that the experimental points fit well to a single exponential decay with a mean lifetime of 130.5 ns is in contrast to most other cases, where metastable lifetimes decays curves investigated in this way represent a mixture or wide distribution of lifetimes \cite{field1993lifetimes}. This may imply that only one single vibration level or a very close group lies just above the barrier to this reaction channel. 

Because significant approximations are involved in the theory behind the analysis of Fig. \ref{fig:lifetime}, the metastable decay was also investigated by a Monte-Carlo type simulation using a realistic model of the apparatus implemented in the software package SIMION \cite{simion}, the results of which are presented in panel B of Fig. \ref{fig:metastable}. The decay was modelled to have a 3 eV kinetic energy release (KER) in line with the observed difference in energy between the thermochemical threshold and the observed appearance energy, reported below. The pink contour shows the simulation results on top of the experimental coincidence map. This simulation fits the experimental observations well with a mean lifetime of 130±10 ns, in good agreement with the simple exponential fit. We note that this lifetime is within the range predicted by Mebel and Bandrauk on the basis of RRKM theory, that is, assuming free flow of internal energy within the molecular ion allowing statistical energy redistribution. 

\subsubsection{Multiple-electron-ion coincidences and action spectra}

\begin{figure}[h]
\centering
\includegraphics[width = 0.5\textwidth]{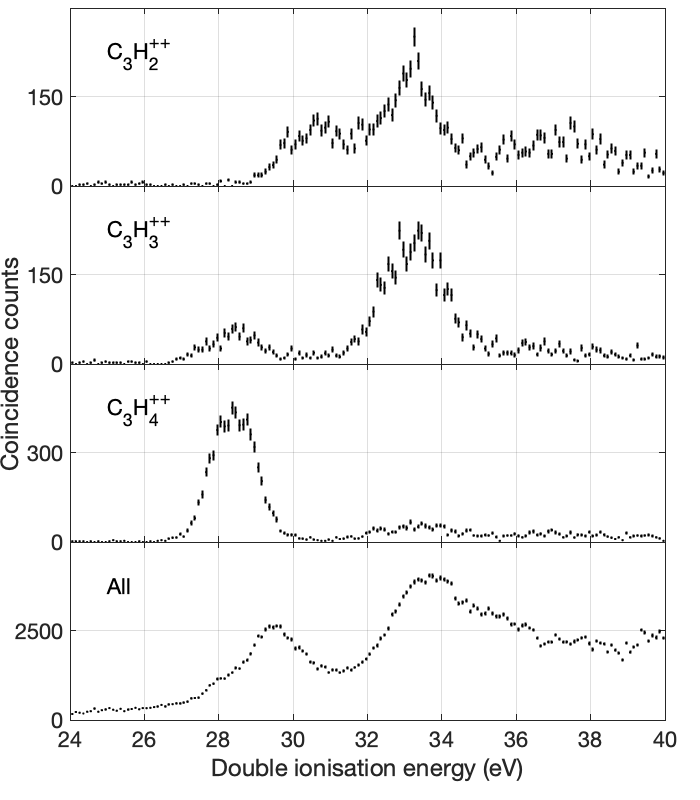}
\caption{\label{fig:DIP_chargeretain} Double ionisation spectra based on electrons that were detected in coincidence with the charge retaining parent species or with charge retaining fragments which lost one or two hydrogen atoms. Because the detected species is doubly ionised and triple ionisation is impossible at this photon energy, no other ions can be involved in these events. 
}
\end{figure}

To determine the appearance energies of the different sets of products and to obtain “action spectra” for the different channels as a function of double ionisation energy, threefold (eei) and fourfold (eeii) coincidences are indispensable. In view of the achievable electron energy resolution, this was primarily done at the photon energy of 40.8 eV. If a particular ion is formed by only one decay channel, an eei spectrum is sufficient to define the action spectrum. This is true for all three doubly charged ions C\subb{3}H\subb{4}\supp{2+}, C\subb{3}H\subb{3}\supp{2+} and C\subb{3}H\subb{2}\supp{2+} for which the spectra are shown in Fig. \ref{fig:DIP_chargeretain}. It is also true for ions formed only in two-body decays of the parent including the metastable decay channel shown in Fig. \ref{fig:metastable}. Analysis of the ion-ion coincidence maps suggests that it applies as a good approximation to H\subb{3}\supp{+} formation (action spectrum in the upper panel of Fig. \ref{fig:DIP_ions}) and to the ionization pathways leading to C\subb{2}H\subb{2}\supp{+} or CH\subb{2}\supp{+} and to C\subb{2}H\subb{3}\supp{+} or CH\supp{+} whose spectra are shown in Fig. \ref{fig:DIP_pi}. The other ions, H\supp{+}, H\subb{2}\supp{+} and C\subb{3}H\subb{n}\supp{+}, are all products of more than one decay channel, so eeii coincidences are formally required to get unambiguous action spectra. Unfortunately, the ion collection efficiency of the apparatus is so low that spectra with useful numbers of counts cannot be obtained for most such fourfold coincidences so eei events have been used to generate the spectra shown in Fig. \ref{fig:DIP_ions}.

From these spectra, we can get an overview of the fates of allene dications in the range of the excitation energies covered by the double photoionization spectrum at 40.8 eV. As can be seen in Fig. \ref{fig:DIP_chargeretain}, the first peak of the total double ionization spectrum is primarily associated with undissociated electronic states of the doubly ionized parent molecule at up to 3 eV above the calculated adiabatic double ionization threshold of 26 eV. Because of an overlap in the mass spectrum of the peaks reflecting the parent ion and the parent ion minus one and minus two hydrogen atoms, it was impossible to separate signals for these species completely. We believe that the peak in the double ionization spectrum of C\subb{3}H\subb{3}\supp{2+} between 26 and 30 eV actually originates from C\subb{3}H\subb{4}\supp{2+} instead, because otherwise its appearance energy would be impossibly low. Also, the high energy part of the C\subb{3}H\subb{4}\supp{2+} spectrum may be affected in a similar way 
In order to demonstrate more realistic spectra for the doubly-charged ions we have carried out an iterative subtraction process, based on the estimated extent of peak overlap.  The resulting spectra are shown in red together with the raw spectra in black.

\begin{figure}[H]
\centering
\includegraphics[width = 0.5\textwidth]{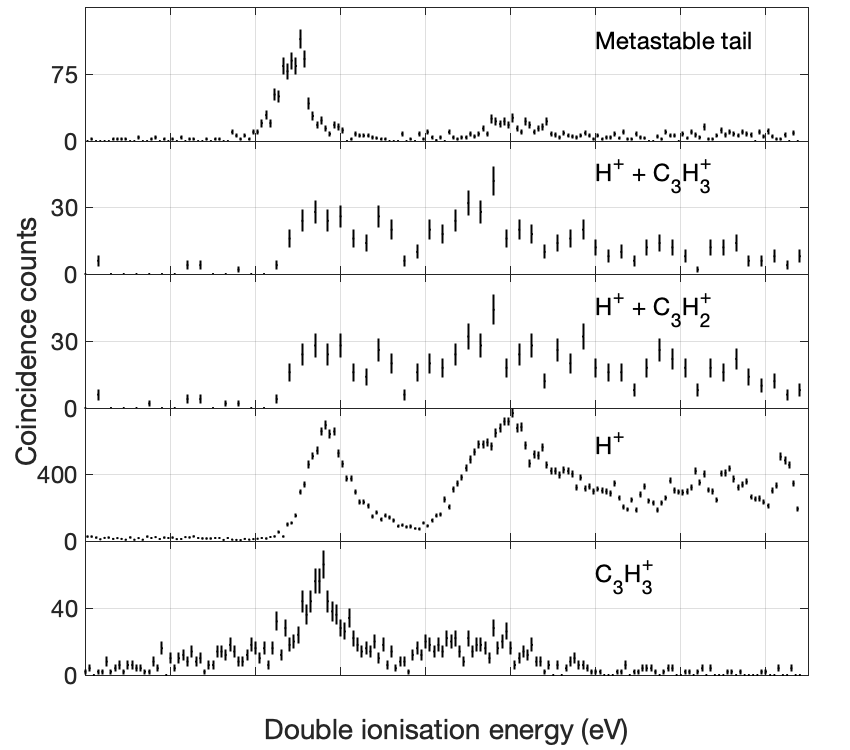}
\caption{\label{fig:DIP_comb} Double ionisation spectra based on electrons that were detected in coincidence with ions as specified in the figure. The uppermost spectrum represents decay events detected in the metastable tail shown in Fig. \ref{fig:metastable}. The next lower spectra are derived from selection of events with two ions such as, $\mathrm{H^+ \;and\; C_3H_3^+}$, and two electrons, eeii. The lower spectra are from events with one ion $\mathrm{H^+ \;or\; C_3H_3^+}$, respectively, and two electrons, eei. One coincident count is equivalent to two electrons being detected in the same event.}
\end{figure}

\begin{figure}
\centering
\includegraphics[width = 0.5\textwidth]{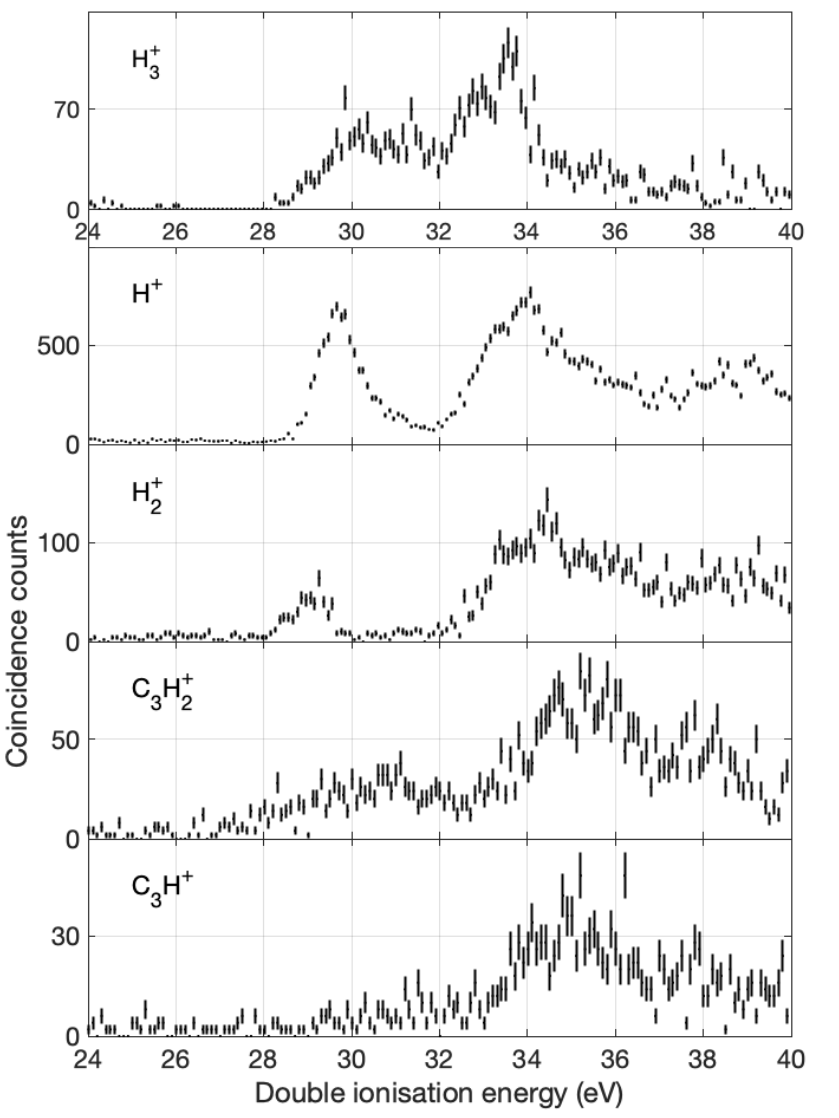}
\caption{\label{fig:DIP_ions} Double ionisation spectra based on two electrons detected in coincidence with a singly charged ion as specified for each spectrum. Both two-body and three-body fragmentations can contribute to these spectra, but two-body decays are probably dominant.  Modify as necessary }
\end{figure}

\begin{figure}
\centering
\includegraphics[width = 0.5\textwidth]{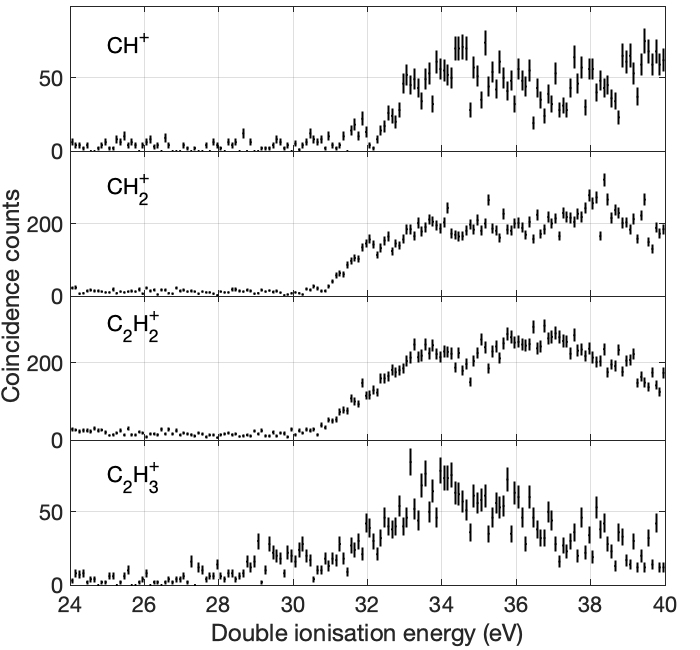}
\caption{\label{fig:DIP_pi} Double ionisation spectra for channels where a C-C bond is broken, based on two electrons detected in coincidence with a singly charged ion as  specified for each spectrum.}
\end{figure}

Over the first 2 eV above the onset of vertical double ionisation at about 27 eV, the parent C$_3$H$_3^{2+}$ ion remains stable on the mass spectrometer time scale. The lowest energy dissociation by charge separation is the slow metastable decay by H\supp{+} ejection, with onset at 28.5 ± 0.3 eV and peak intensity at 29 eV $\pm$ 0.3 eV shown in Fig. \ref{fig:DIP_comb}. The metastable signal appears only in a narrow range, presumably just above threshold, \textcolor{blue}{with a width of ...}. in line with the expected electron energy resolution at this ionisation energy of ca 0.6 eV. This observation strengthens the view that a single vibrational level of the parent dication is responsible for the slow dissociation. Once the ionisation energy exceeds 29 eV formation of H$^+$ + C$_3$H$_3^+$ occurs rapidly and becomes the most probable dissociation pathway. But at the same energy both H$_2^+$ and H$_3^+$ are formed with low intensity (cf. Fig. \ref{fig:DIP_ions}). The three lowest panels in Fig. \ref{fig:DIP_comb} all represent the process of H$^+$ + C$_3$H$_3^+$ formation detected in different ways, and the differences between them illustrate the problems of poor statistics in fourfold coincidences and background subtraction of overlapping mass peaks.

Fig. \ref{fig:DIP_ions} shows the yields of selected single ions in coincidence with electron pairs. For all the C$_3$H$_n^+$ ions there is an overlap problem similar to that encountered for the doubly-charged ions (cf. Fig. \ref{fig:DIP_chargeretain}) and we have compensated for it in a similar way.  In this case, fourfold coincidence measurements give clear guidance on the extent of overlap and the necessary subtractions.  Because of the threefold coincidence selection, the contributing ions may be formed by three-body as well as two-body dissociations, but the lowest energy pathways must be the two-body reactions forming H$^+$, H$_2^+$ and H$_3^+$ with observed onset energies in the region of 28.5 to 29.5 eV. Where two-body reactions dominate, the spectra for H$_3^+$ and C$_3$H$^+$ and those of H$_2^+$ and C$_3$H$_2^+$ should be the same.  This is borne out by comparison of the spectra up to 35 or 36 eV, but at higher energy the different intensities indicate that three-body reaction releasing additional neutral fragments take over. The calculations of Mebel and Bandrauk\cite{mebel2008theoretical} indicate that barriers to the three hydrogen ion loss pathways should lie in the range of 28.84 to 29.03 eV, and their RRKM calculations predict that H\subb{3}\supp{+} peak formation should occur at 29.53 eV. The calculated onsets and the peak production energy agree well with the observations in Fig. \ref{fig:DIP_ions}. However, further predictions based on the RRKM model assumption that all channels are in competition at all energies are not in line with our experimental data since the intensities of the channels leading to H$^+$ and H$_3^+$ are not comparable at any energy. An alternative interpretation, that chimes with the “roaming” mechanism is that some of the initial population becomes isolated as a $[\mathrm{C_3H_2-H_2}]^{++}$ complex, where the $\mathrm{H_2}$ can either escape or capture a proton. Such a mechanism would explain why $\mathrm{H_3^+}$ formation does not take over from $\mathrm{H^+}$ production and why the intensities and spectra for H$^+$ and H$_3^+$ production are similar.

In the higher double ionization energy range, one of the main observation from our data is that the three-body products $\mathrm{H^+ + C_3H_2^+ + H}$ take over completely from $\mathrm{H^+ + C_3H_3^+}$ at energies above 34 eV, as demonstrated by the spectra in Fig. \ref{fig:DIP_comb}. The most natural explanation for that is that this is a sequential decay; $\mathrm{C_3H_3^+}$ ions with sufficient internal energy (8 eV initially) to dissociate further by H atom loss. Another very interesting channel is the production of $\mathrm{H_2^+ + C_3H_2^+}$, which according to the eei data given in Fig. \ref{fig:DIP_ions} occurs weakly between 28 and 30 eV, then more strongly above 32 eV. Formation near 29 eV is possible only for the cyclic form of C$_3$H$_2^+$ (thermochemical threshold 25.6 eV) with a sensible kinetic energy release. Both the cyclic and perhaps more naturally produced linear C$_3$H$_2^+$ ion (threshold 27.8 eV) can be formed at 32 eV, which is the energy where the second main double ionization band begins (cf. Fig. \ref{fig:DIP}). According to a recent fs laser pump-probe study of the roaming mechanism for $\mathrm{H_3^+}$ formation from doubly ionised methanol \cite{livshits2020time}, the two exit channels giving $\mathrm{H_3^+}$ and $\mathrm{H_2^+}$ are both ultrafast and in competition. If that is also true here, the channel forming H$_3^+$ must have entropic or related factors in its favour, as it takes over completely from H$_2^+$ formation in the second part of the first main double ionisation band (cf. Fig. \ref{fig:DIP_ions}). It is also striking that the charge retaining channel $\mathrm{H_2 + C_3H_2^{2+}}$ appears at 29.6 ± 0.3 eV (cf. Fig. \ref{fig:DIP_chargeretain}), the energy where H$_2^+$ ceases to be produced and close to its calculated asymptote. This suggests that charge separation and charge retention by the heavier fragment are also in competition at this point.

There is an apparent reappearance of intensity for H$^+$ + C$_3$H$_3^+$ peaking at around 34 eV (cf. Fig. \ref{fig:DIP_comb}) in the second double ionisation band. In this range both the cyclic and linear isomers of the ion may be formed and charge-retaining formation of both C$_3$H$_3^{2+}$ and C$_3$H$_2^{2+}$ appear to compete strongly in the same energy range.  As can be seen from Fig. \ref{fig:DIP_pi}, this is also the energy range where the charge-separating C-C bond-breaking reaction giving rise to $\mathrm{CH_2^+ + C_2H_2^+}$ becomes intense. It is slightly surprising that so many very different dissociation routes can all compete in the same energy range with comparable intensities in all the channels.

\begin{table}[H]
    \centering
    \caption{Double ionisation of allene at 40.8 eV photon energy in comparison to thermodynamic thresholds and theoretical predictions. The uncertainties in the observed values are a consequence of the kinetic energy resolution of the electron spectrometer used.}
\begin{tabular}{|c|cc|cc|}
    \hline
    \multirow{2}{*}{Channel} & \multicolumn{2}{c|}{0 K threshold (eV)} & \multicolumn{2}{c|}{Appearance energy (eV)} \\
    & Thermo. & Calc (M\&B)  &  Calc (M\&B) & Obs. (us) \\
    \hline
    \multirow{2}{*}{H\supp{+} + 39\supp{+}} & cyclic 25.0  & 24.95  & 28.8 &  28.9 $\pm$0.3  \\
     & linear 26.1 & 26.2 & & \\
     \hline
    \multirow{2}{*}{H\supp{+} + 38\supp{+} + H }& cyclic 28.3  &  & & 32$\pm$0.3 \\
     & linear 30.4 & & & \\
     \hline
     \multirow{2}{*}{H\supp{+} + 37\supp{+} + H\subb{2}} & 30.4 &  & &  \\
    & & & &\\
    \hline
    \multirow{2}{*}{H\subb{2}\supp{+} + 38\supp{+} }& cyclic 25.6 & 27.7 & & 32.8$\pm$0.2\\
    & linear 27.8 & 27.9 & 29.44 &   \\
    \hline
    \multirow{2}{*}{H\subb{2}\supp{+} + 37\supp{+} + H} & 32.2 &   & & \\
    & & & &\\
    \hline
     \multirow{2}{*}{H\subb{3}\supp{+} + 37\supp{+}} &26.0 & 26.0    & 29.03 & 29$\pm$0.3\\
     & & & &\\
    \hline
     \multirow{2}{*}{CH\subb{2}\supp{+} + C\subb{2}H\subb{2}\supp{+}} & 26.1 & 26.2 & 30.8 (vinylidene) & 31$\pm$0.3\\
     & & & &\\
    \hline
     \multirow{2}{*}{CH\supp{+} + C\subb{2}H\subb{3}\supp{+} }& & &30.36 & 32.1$\pm$0.2 \\
     & & & &\\
    \hline
    C\subb{3}H\subb{4}\supp{2+} & & \makecell{adb. 25.84 \\ vert. 28.05} & & \makecell{ \\ 27.5$\pm$0.3 }\\
    \hline
     \multirow{2}{*}{C\subb{3}H\subb{3}\supp{2+} + H }& & c-30.55 & & 32.1$\pm$0.2\\
     & & & &\\
    \hline
     \multirow{2}{*}{C\subb{3}H\subb{2}\supp{2+} + H\subb{2}} & & 29.92, 29.39 & &29.6$\pm$0.3\\
     & & & &\\
    \hline
\end{tabular}
    \label{tab:app_energy}
\end{table}

The appearance energies of the different sets of products determined from the experimental spectra in Figs. \ref{fig:DIP_chargeretain}-\ref{fig:DIP_pi}, are listed in Table \ref{tab:app_energy} together with thermodynamic thresholds and theoretical predictions. For all the thermodynamic 0K thresholds for ions formed with no internal energy or kinetic energy, heats of formation are taken from the NIST database or are estimated by combining known thermodynamic data with theoretical calculations (e.g. from Mebel and Bandrauk (M\&B)\cite{mebel2008theoretical}). We note that the ions of mass 39, 38 and possibly 37 can have either cyclic or linear structures, with significantly different heats of formation (the heats of formation for mass 37 is uncertain). The Mebel and Bandrauk-calculated thresholds are taken from their figure 2, with some degree of ambiguity.

\section{Conclusions}

Using multi-particle coincidence experiments we have obtained single-photon double and triple ionization spectra of allene, and have determined how dissociation of the doubly charged ions depends on the ionisation energy.  New high-level calculations confirm that adiabatic double ionisation would require isomerization to a different structure, not accessible to vertical ionisation processes.  The calculations allow substructure in the first double ionisation band to be assigned to different electronic states of the dications. Double ionisation of allene by Auger decay of C1s vacancy states is found to populate different spectra of dication states according to location of the vacancy on either the central or an outer C atom.

Triple ionization of allene by three routes, valence ionization at 100 eV, double Auger decay of C1s vacancies and Auger decay of a C1s core-valence doubly ionised intermediate state yield different spectra, but all exhibit an onset of triple ionization at approximately 50 eV, in line with predictions available in the literature. 

Eight significant decay pathways for dissociative decay of nascent allene dications have been identified and the energy dependence of their relative intensities is reported.  At the lowest energies there is evidence that formation of H$_2^+$ + C$_3$H$_2^+$ and H$_3^+$ + C$_3$H$^+$ are in competition with each other and possibly with the charge-retaining H + C$_3$H$_3^{2+}$ channel, but not with charge separation to H$^+$ + c-C$_3$H$_3^+$. This finding supports the roaming mechanism, already proposed, where an H$_2$ molecule becomes partially detached from the heavy residual molecular dication.   
	For the decay to H$^+$ + c-C$_3$H$_3^+$, we have confirmed the existence of a slow metastable decay happening in a narrow energy range just above threshold. This decay is well characterised as a single exponential with a mean lifetime determined as 130 ±10 ns.  We suspect that a single vibrational level of the parent dication may be involved and might be identified by future calculations.

\begin{acknowledgments}
This work has been financially supported by the Swedish Research Council (VR) and the Knut and Alice Wallenberg Foundation, Sweden. We thank the Helmholtz Zentrum Berlin for the allocation of synchrotron radiation beam time and the staff of BESSY-II for smooth running of the storage ring during the single-bunch runtime. The research leading to these results has received funding from the European Union’s Horizon 2020 research and innovation programme under grant agreement No 730872.
\end{acknowledgments}

\appendix

\bibliography{references}

\end{document}